\documentclass[a4paper,11pt]{article}
\usepackage{jheppub}

\makeatletter
\def\@fpheader{\relax}
\makeatother

\begin{document}

\title{Remnants of Black Rings from Gravity's Rainbow}

\author[a,b]{Ahmed Farag Ali}
\author[c]{Mir Faizal}
\author[d]{Mohammed M. Khalil}

\affiliation[a]{Center for Fundamental Physics, Zewail City of Science and Technology,\\Giza 12588, Egypt}
\affiliation[b]{Deptartment of Physics, Faculty of Science, Benha University,\\Benha 13518, Egypt}
\affiliation[c]{Department of Physics and Astronomy, University of Waterloo,\\Waterloo, Ontario, N2L 3G1, Canada}
\affiliation[d]{Department of Electrical Engineering, Alexandria University,\\ Alexandria 12544, Egypt}

\emailAdd{afarag@zewailcity.edu; ahmed.ali@fsc.bu.edu.eg}
\emailAdd{f2mir@uwaterloo.ca}
\emailAdd{moh.m.khalil@gmail.com}

\abstract{In this paper, we  investigate a spinning black ring and a charged black ring in the context of gravity's rainbow. By incorporating rainbow functions proposed by Amelino-Camelia, et al. in \cite{Amelino1996pj,amelino2013} in the metric of the black rings, a considerable modification happens to their thermodynamical properties. We calculate corrections to the temperature, entropy and heat capacity of the black rings. These calculations demonstrate that the behavior of Hawking radiation changes considerably near the Planck scale in  gravity's rainbow, where it is shown that black rings do not evaporate completely and a remnant is left as the black rings evaporate down to Planck scale.}


\arxivnumber{}

\maketitle
\flushbottom

\section{Introduction}

Developing a quantum theory of gravity is currently one of the main challenges in theoretical physics. Various approaches to quantum gravity suggest a departure from the standard energy-momentum dispersion relation, $E^2-p^2=m^2$.
These approaches include spacetime discreteness \cite{'tHooft:1996uc}, spontaneous symmetry breaking of Lorentz invariance in string field theory \cite{Kostelecky:1988zi}, spacetime foam models \cite{Amelino1997gz} and spin-network in loop quantum gravity (LQG) \cite{Gambini:1998it}. Besides,
there are other approaches such as non-commutative geometry \cite{Carroll:2001ws} which predicts a Lorentz invariance violation at the Planck scale. Recently, a new approach to quantum gravity was formulated by Horava in \cite{Horava:2009uw,Horava:2009if} that predicts a modification of the dispersion relation at the scale of quantum gravity.
All these indications suggest that Lorentz violation or deformation may be an essential property in constructing a quantum theory of gravity. This departure can be collectively expressed in the following form:
\begin{equation}
E^2f^2(E/E_P)-p^2g^2(E/E_P)=m^2
\end{equation}
where $E_P$ is the Planck energy, and the functions $f(E/E_P)$ and $g(E/E_P)$ satisfy
\begin{equation}
\lim\limits_{E/E_P\to0} f(E/E_P)=1,\qquad \lim\limits_{E/E_P\to0} g(E/E_P)=1.
\end{equation}

One of the interesting approaches that naturally produce modified dispersion relations (MDR) is called doubly special relativity (DSR) \cite{AmelinoCamelia:2000mn}. DSR can be briefly described as an extension to special relativity that includes an invariant energy scale \cite{Magueijo:2001cr}, usually assumed to be the Planck energy, in addition to the invariance of the speed of light. There are other reasons to deform the dispersion relation. In string theory, it is not possible to probe spacetime below the string length scale. Thus, string theory comes naturally equipped with a minimum length scale, which is the string length scale \cite{Amati:1988tn,Garay:1994en}. A minimum length scale in string theory could also be translated to a maximum energy scale. This maximum energy scale will in turn deform the dispersion relation to the one used in DSR \cite{Ali:2009zq,Ali:2011fa}.

In 2004, Magueijo and Smolin proposed \cite{Magueijo:2002xx} an extension to DSR to include curvature i.e. doubly general relativity. In this approach, the geometry of spacetime depends on the energy $E$ of the particle used to probe it. Thus, spacetime is represented by a one parameter family of metrics parametrized by the ratio $E/E_P$, forming a \emph{rainbow} of metrics, and hence the name \emph{gravity's rainbow}.
The modified metric in gravity's rainbow is evaluated from \cite{Magueijo:2002xx}
\begin{equation}
\label{rainmetric}
g(E)=\eta^{ab}e_a(E)\otimes e_b(E),
\end{equation}
and the energy dependence of the frame fields is given by
\begin{equation}
e_0(E)=\frac{1}{f(E/E_P)}\tilde{e}_0, \qquad
e_i(E)=\frac{1}{g(E/E_P)}\tilde{e}_i,
\end{equation}
where the tilde quantities refer to the energy independent frame fields. Therefore, the metric in flat spacetime takes the form\cite{Magueijo:2002xx}
\begin{equation}
ds^2=-\frac{dt^2}{f(E)^2}+\frac{dx_i dx^i}{g(E)^2},
\end{equation}
and the modified Schwarzschild metric \cite{Magueijo:2002xx}
\begin{equation}
ds^2=-\frac{1}{f(E)^2}\left(1-\frac{2M}{r}\right)dt^2+\frac{1}{g(E)^2}\left( \frac{dr^2}{1-\frac{2M}{r}}+{r^2}d\Omega^2 \right),
\end{equation}
where we use units in which $G=c=\hbar=k_B=1$.

From Eq.\eqref{rainmetric} and the previous examples, we see that to get the modified metric in gravity's rainbow, we simply change $dt\to dt/f(E/E_P)$ and all spatial coordinates $dx^i\to dx^i/g(E/E_P)$. 

The choice of the rainbow functions $f(E/E_P)$ and $g(E/E_P)$ is important for making predictions. This choice is supposed to be based on phenomenological motivations. Among different arbitrary choices in \cite{Garattini:2011hy,Leiva:2008fd,Li:2008gs,Ali:2014cpa,Ali:2014xqa,Awad:2013nxa,Barrow:2013gia,Liu:2007fk}, many aspects of the theory have been studied with FRW universe and black hole thermodynamics. Among the proposals in the literature, we use one of the most interesting and the most studied modified dispersion relations that was proposed by Amelino-Camelia, et al. in \cite{Amelino1996pj,amelino2013}
\begin{equation}
\label{rainbowfns}
f\left(E/{E_P}\right)=1,\qquad g\left( E/{E_P} \right)=\sqrt{1-\eta \left(\frac{E}{E_P}\right)^{n}},
\end{equation}
These functions are compatible with results from loop quantum gravity and $\kappa$-Minkowski non-commutative spacetime\cite{amelino2013}. For a discussion about the
phenomenological implications of Eq. (\ref{rainbowfns}), it is very useful to consult the discussion in the detailed review \cite{amelino2013}.

In this paper, we extend the study of black holes in gravity's rainbow by one of authors in \cite{Ali:2014xqa} but this time for  the thermodynamics of black rings in the framework of gravity's rainbow. We find that the black ring reaches a remnant at which the temperature, entropy, and heat capacity go to zero. This result is similar to the one in Ref. \cite{Ali:2014xqa} for black hole remnants in gravity's rainbow. 

\section{Spinning Black Rings}
In this section, we briefly review the thermodynamical properties of spinning black rings. The metric of a spinning neutral black ring in five dimensions takes the form \cite{Emparan:2001wn,Zhao:2006zw,Altamirano:2014tva}
\begin{eqnarray}
\label{ringmetric}
ds^2=&&-\frac{F(y)}{F(x)}\left(dt+C(\nu,\lambda)R\frac{1+y}{F(y)}d\psi\right)^2 \nonumber\\
&& +\frac{R^2F(x)}{(x-y)^2}\left(\frac{dx^2}{G(x)}+\frac{G(x)}{F(x)}d\phi^2-\frac{G(y)}{F(y)}d\psi^2-\frac{dy^2}{G(y)}\right)
\end{eqnarray}
where
\begin{eqnarray}
& F(\xi)=1+\lambda\xi,\qquad G(\xi)=(1-\xi^2)(1+\nu\xi), \nonumber\\
& C(\nu,\lambda)=\sqrt{\lambda(\lambda-\nu)\frac{1+\lambda}{1-\lambda}}.
\end{eqnarray}
The dimensionless parameters $\lambda$ and $\nu$ take values in the range $0<\nu\leq\lambda<1$, and to avoid conical singularity at $x=1$ they must be related by
\begin{equation}
\lambda=\frac{2\nu}{1+\nu^2}.
\end{equation}
To avoid conical singularities at $x=-1$ and $y=-1$, the angular coordinates $\phi$ and $\psi$ are chosen to have periodicity
\begin{equation}
\Delta\phi=\Delta\psi=2\pi\frac{\sqrt{1-\lambda}}{1-\nu}.
\end{equation}
The coordinates $x,y$ are restricted to the ranges $-1\leq x\leq 1$ and $-1/\nu\leq y<-1$. The event horizon is located at $y_h=-1/\nu$.

To find the temperature of the black ring, we can use the formula derived in \cite{Zhao:2006zw}, which applies to all types of black rings,
\begin{equation}
\label{temp}
T=\frac{1}{4\pi}\sqrt{A_{,y}(x,y_h)B_{,y}(x,y_h)}
\end{equation}
where the functions $A(x,y)$ and $B(x,y)$ are found by expressing the black ring metric in the following form
\begin{equation}
\label{metric}
ds^2=-Adt^2+\frac{1}{B}dy^2+g_{\psi\psi}(d\psi+N^\psi dt)^2+g_{xx}dx^2+g_{\phi\phi}d\phi^2.
\end{equation}
For the metric in Eq. \eqref{ringmetric},
\begin{eqnarray} &&A(x,y)=\frac{F(y)}{F(x)}\left(1-\frac{C(\nu,\lambda)^2(1+y)^2}{\frac{1}{(x-y)^2}F(x)^2G(y)+C(\nu,\lambda)^2(1+y)^2}\right) \nonumber\\
&&B(x,y)=-\left(\frac{R^2}{(x-y)^2}\frac{F(x)}{G(y)}\right)^{-1}.
\end{eqnarray}
Thus, from Eq.\eqref{temp} the temperature is \cite{Zhao:2006zw,Altamirano:2014tva}
\begin{equation}
\label{temp1}
T=\frac{(1-\nu)\sqrt{1+\nu^2}}{4\sqrt{2}\pi\nu R}.
\end{equation}

The entropy can be calculated from the first law of black hole thermodynamics \cite{Emparan:2004wy,Altamirano:2014tva}
\begin{equation}
\label{1stlaw}
dS=\frac{1}{T}dM-\frac{\Omega}{T}dJ-\frac{\Phi}{T}dQ
\end{equation}
where the mass $M$, angular velocity $\Omega$, angular momentum $J$, electrostatic potential $\Phi$, and charge $Q$ are given by \cite{Emparan:2006mm, Altamirano:2014tva}
\begin{eqnarray}
\label{parameters}
&&M=\frac{3\pi R^2\nu}{2(1-\nu)(1+\nu^2)}, \qquad \Omega=\frac{1}{R}\sqrt{\frac{1-\nu+\nu^2-\nu^3}{2+2\nu}}, \nonumber\\
&&J=\frac{\pi\nu R^3}{\sqrt{2}}\left(\frac{1+\nu}{(1-\nu)(1+\nu^2)}\right), \qquad \Phi=Q=0.
\end{eqnarray}
Substituting those relations in Eq.\eqref{1stlaw}, we get the entropy
\begin{equation}
\label{entropy1}
S=\frac{2\sqrt{2}\pi^2R^3\nu^2}{(1-\nu)(1+\nu^2)^{3/2}}.
\end{equation}

The thermodynamic stability of black holes and black rings is determined by the heat capacity at constant angular momentum $C_J$ \cite{Monteiro:2009tc}, which can be calculated from the thermodynamic relation
\begin{equation}
\label{heatcap}
C_J=T\left(\frac{\partial S}{\partial T}\right)_J
\end{equation}
The partial derivative of entropy with respect to temperature keeping the angular momentum fixed is given by the identity \cite{thorade2013partial}
\begin{equation}
\left(\frac{\partial S}{\partial T}\right)_J=\frac{\left(\frac{\partial S}{\partial R}\right)_\nu \left(\frac{\partial J}{\partial \nu}\right)_R- \left(\frac{\partial S}{\partial \nu}\right)_R \left(\frac{\partial J}{\partial R}\right)_\nu}{\left(\frac{\partial T}{\partial R}\right)_\nu \left(\frac{\partial J}{\partial \nu}\right)_R - \left(\frac{\partial T}{\partial \nu}\right)_R \left(\frac{\partial J}{\partial R}\right)_\nu}
\end{equation}
which leads to the relation \cite{Monteiro:2009tc,Altamirano:2014tva}
\begin{equation}
\label{capacity1}
C_J=\frac{12\sqrt{2}\pi^2R^3\nu^2(\nu-1/2)\sqrt{1+\nu^2}}{(1-\nu)(2+\nu^2)(1+\nu^2)^2}.
\end{equation}
We see that when $\nu>1/2$ the heat capacity is positive, which means the black ring is thermodynamically stable. However, when $\nu<1/2$ the heat capacity is negative and the black ring is thermodynamically unstable.

From the previous relations, we see that as $\nu\to0$, the temperature goes to infinity while the mass, angular momentum, entropy, and heat capacity go to zero. This implies that the black ring \emph{evaporates} as $\nu\to0$.

\section{Spinning Black Rings in Gravity's Rainbow}
To get the modified metric of black rings in gravity's rainbow, change $dt\to dt/f(E/E_P)$ and all spatial coordinates $dx^i\to dx^i/g(E/E_P)$. The metric \eqref{metric} becomes
\begin{equation}
ds^2=-\frac{A}{f^2(E)}dt^2+\frac{1}{Bg^2(E)}dy^2+g_{\psi\psi}\left(\frac{d\psi}{g(E)}+\frac{N^\psi}{f(E)}dt\right)^2+\frac{g_{xx}}{g^2(E)}dx^2+\frac{g_{\phi\phi}}{g^2(E)}d\phi^2.
\end{equation}
Thus, the modified temperature can be calculated from Eq.\eqref{temp} with the change $A(x,y)\to A(x,y)/f(E)^2$ and $B(x,y)\to B(x,y)g(E)^2$
\begin{equation}
\label{modtemp}
T'=\frac{1}{4\pi}\sqrt{A_{,y}(x,y_h)B_{,y}(x,y_h)}\frac{g(E)}{f(E)}.
\end{equation}
Using the rainbow functions from Eq. \eqref{rainbowfns} we get
\begin{equation}
T'=\frac{(1-\nu)\sqrt{1+\nu^2}}{4\sqrt{2}\pi\nu R}\sqrt{1-\eta\left(\frac{E}{E_P}\right)^n}.
\end{equation}

According to \cite{Adler:2001vs, Cavaglia:2003qk, Medved:2004yu, AmelinoCamelia:2004xx}, the uncertainty principle $\Delta p\geq 1/\Delta x$ can be translated to a lower bound on the energy $E\geq 1/\Delta x$ of a particle emitted in Hawking radiation, and the value of the uncertainty in position can be taken to be the event horizon radius. For a black ring, the dimensionless parameter $\nu$ determines the shape of the horizon, and can be considered as a measure for the radius of the ring, i.e. small $\nu$ corresponds to a thin ring \cite{Emparan:2001wn}. Thus, the value of $\Delta x$ is proportional to $\nu$, and we need to multiply it by the scale factor $R$ to have length units. Hence,
\begin{equation}
E\geq \frac{1}{\Delta x} \approx \frac{1}{\nu R}.
\end{equation}
The temperature becomes
\begin{equation}
\label{modtemp1}
T'=\frac{(1-\nu)\sqrt{1+\nu^2}}{4\sqrt{2}\pi\nu R}\sqrt{1-\eta\left(\frac{1}{\nu RE_P}\right)^n}.
\end{equation}

\begin{figure}[h]
\centering
\includegraphics[width=0.55\linewidth]{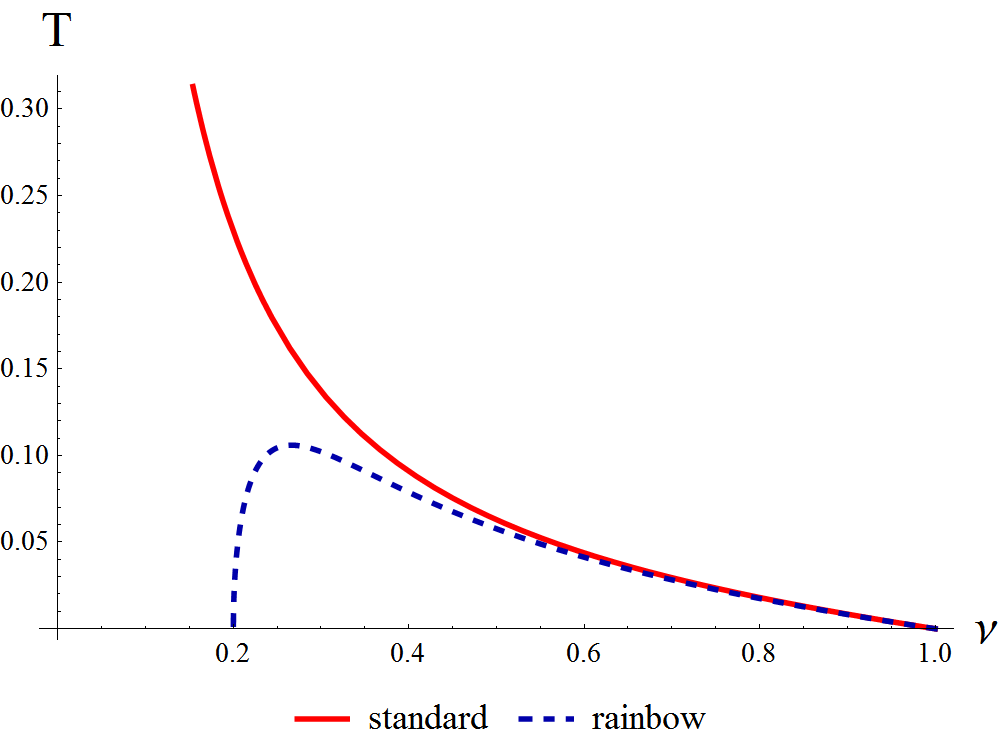}
\caption{\label{fig:temp} Temperature of spinning black rings as a function of $\nu$. In gravity's rainbow, the temperature goes to zero when $\nu\to\eta^{\frac{1}{n}}/RE_P$. To plot Eq.\eqref{modtemp1}, we used the generic values $E_P=5, n=2, \eta=1,$ and $R=1$.}
\end{figure}

From Eq. \eqref{modtemp1}, it is clear that the temperature goes to zero at $\nu=\eta^{\frac{1}{n}}/RE_P$. Substituting this into Eq.\eqref{parameters} we get the minimum mass
\begin{equation}
M_{min}=\frac{3\pi R^4E_P^2\eta^{1/n}}{2\left(RE_P-\eta^{1/n}\right)\left(R^2E_P^2+\eta^{2/n}\right)}.
\end{equation}
This implies that the black ring has a remnant. Fig.\ref{fig:temp} is a plot of the modified temperature with the generic values $E_P=5, n=2, \eta=1,$ and $R=1$; different values lead to the same qualitative behavior. This modification is generated by the choice of the rainbow functions \eqref{rainbowfns}. Other functions have different effects. For example, if $f(E)=g(E)$, as in the case of the MDR proposed in \cite{Magueijo:2002xx}, there is no modification to the thermodynamics. However, if $f(E)\approx 1+\alpha E/E_P, g(E)=1$, as in \cite{Amelino1997gz}, the temperature is modified but there is no remnant.

\begin{figure}[h]
\centering
\includegraphics[width=0.55\linewidth]{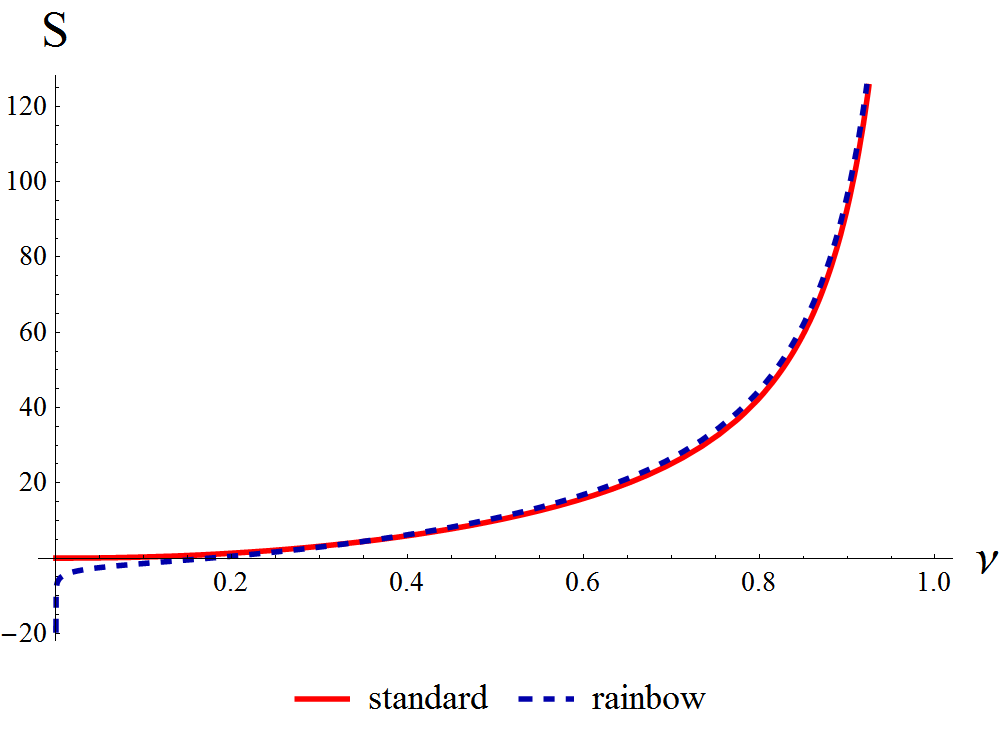}
\caption{\label{fig:ent} Entropy of spinning black rings as a function of $\nu$. The entropy goes to zero as $\nu\to\eta^{\frac{1}{n}}/RE_P$ and stops there; the negative value of the entropy is because of the approximation we made to do the integral \eqref{dSmod}.}
\end{figure}

The modified entropy can be calculated from the first law of black hole thermodynamics \eqref{1stlaw} using the parameters in Eq. \eqref{parameters} and the modified temperature \eqref{modtemp1}
\begin{equation}
\label{dSmod}
dS'=\frac{2\sqrt{2}\pi^2R^3\nu(2\nu^3-\nu^2-\nu+2)}{(\nu-1)^2(\nu^2+1)^{5/2}\sqrt{1-\eta\left(\frac{1}{\nu R E_P}\right)^n}}d\nu
\end{equation}
This equation cannot be integrated even for specific values of $n$. Expanding to first order in $\eta$ and integrating for $n=2$ we get
\begin{eqnarray}
S'=&&\frac{2\sqrt{2}\pi^2R^3\nu^2}{(1-\nu)(1+\nu^2)^{3/2}}+\frac{\sqrt{2}\pi^2 R\eta}{E_P^2} \nonumber\\
&&\left(\frac{\nu^4-2}{(\nu-1)(1+\nu^2)^{3/2}}+\frac{1}{\sqrt{2}}\ln\left[\frac{1+\nu+\sqrt{2+2\nu^2}}{1-\nu}\right]+2\ln\left[\frac{\nu}{1+\sqrt{1+\nu^2}}\right]\right)
\end{eqnarray}
We plotted this result in Fig.\ref{fig:ent}. From the plot, we see that the entropy becomes negative near $\nu=0$, but this is because of the approximation we made to evaluate the integral. The derivative $dS/d\nu$, however, goes to infinity at $\nu=\eta^{\frac{1}{n}}/RE_P$, which means the entropy is not continuous at this value; the entropy goes to zero.

\begin{figure}[h]
\centering
\includegraphics[width=0.55\linewidth]{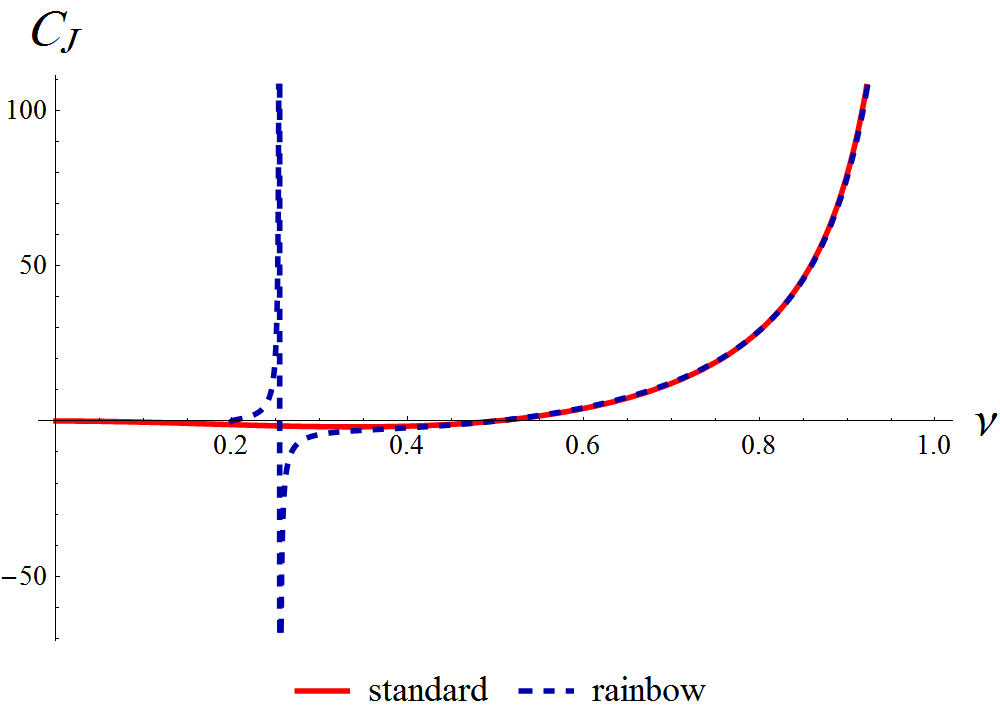}
\caption{\label{fig:capacity} Heat capacity of spinning black rings as a function of $\nu$. The heat capacity diverges at the value of maximum temperature and then goes to zero as $\nu\to\eta^{\frac{1}{n}}/RE_P$}
\end{figure}

To get the modified heat capacity, we use the modified temperature with $n=2$ and the modified entropy in Eq.\eqref{heatcap} to get
\begin{equation}
C'_J=\frac{12\sqrt{2}\pi^2E_P^2R^5\nu^4(\nu-1/2) \sqrt{1-\eta/(\nu RE_P)^2}}{(1-\nu)\sqrt{1+\nu^2}\left(E_P^2R^2\nu^2(\nu^4+3\nu^2+2)+\eta(4\nu^4+3\nu^3-6\nu^2+3\nu-4)\right)}
\end{equation}
which leads to the standard result \eqref{capacity1} when $\eta\to 0$. Fig.\ref{fig:capacity} is a plot of this relation, and we see that it diverges at a point at which the temperature reaches its maximum value, and then goes to zero at $\nu=\eta^{\frac{1}{n}}/RE_P$. The zero value of the heat capacity means the black ring cannot exchange heat with the surrounding space, and hence predicting the existence of a remnant.

\section{Charged Black Rings}
The metric of a static electrically charged dilatonic black ring takes the form \cite{Kunduri:2004da, Elvang:2003mj,Grunau:2014vwa}
\begin{eqnarray}
ds^2=&&-\frac{F(x)}{F(y)}\frac{dt^2}{V_\beta(x,y)^{2/3}}+\frac{R^2V_\beta(x,y)^{1/3}}{(x-y)^2} \nonumber\\
&&\left[-F(x)\left((1-y^2)d\psi^2+\frac{F(y)}{1-y^2}dy^2\right)+F(y)^2\left(\frac{dx^2}{1-x^2}+\frac{1-x^2}{F(x)}d\phi^2\right)\right]
\end{eqnarray}
where
\begin{equation}
F(\xi)=1-\lambda\xi, \qquad V_\beta(x,y)=\cosh^2\beta-\frac{F(x)}{F(y)}\sinh^2\beta
\end{equation}
The dimensionless parameter $\lambda$ is in the range $0<\lambda<1$, and the coordinate ranges are $-1\leq x\leq 1$, $-\infty<y\leq -1$ and $1/\lambda<y<\infty$. The horizon is located at $y_h=-\infty$.

To calculate the temperature we use Eq.\eqref{temp} with
\begin{eqnarray}
&& A(x,y)=\frac{F(x)}{F(y)}\frac{1}{V_\beta(x,y)^{2/3}} \nonumber\\
&& B(x,y)=\frac{(x-y)^2(1-y^2)}{R^2V_\beta(x,y)^{1/3}F(x)F(y)}
\end{eqnarray}
leading to \cite{Kunduri:2004da, Zhao:2006zw}
\begin{equation}
T=\frac{1}{4\pi R\lambda\cosh\beta}
\end{equation}

To find the entropy we use the first law of black hole thermodynamics \eqref{1stlaw} with the parameters \cite{Kunduri:2004da,Grunau:2014vwa}
\begin{eqnarray}
&&M=\frac{3}{4}\pi R^2\lambda(1+\lambda)\left(1+\frac{2}{3}\sinh^2\beta\right), \qquad \Omega=J=0, \nonumber\\
&&Q=\pi R^2\lambda(1+\lambda)\cosh\beta\sinh\beta, \qquad \Phi=\tanh\beta
\end{eqnarray}
which lead to the entropy
\begin{equation}
S=\pi^2R^3\lambda^2\left(\frac{1}{2}+\frac{2}{3}\lambda\right)\cosh(\beta)\left[4-\cosh(2\beta)\right].
\end{equation}
The heat capacity can be calculated via Eq. \eqref{heatcap}, and since $J=0$, we can simply evaluate it by differentiating the temperature and entropy with respect to $\lambda$
\begin{equation}
C=T\frac{\partial S}{\partial T}=T\frac{\partial S/\partial \lambda}{\partial T/\partial \lambda}= \pi^2R^3\lambda^2(1+2\lambda)\cosh(\beta)\left[\cosh(2\beta)-4\right]
\end{equation}

As for the case of spinning black ring, the temperature of a charged black ring goes to infinity as $\lambda\to 0$, while the mass, entropy, and heat capacity go to zero. This implies that the charged black ring \emph{evaporates} as $\lambda\to0$.

\section{Charged Black Rings in Gravity's Rainbow}
In gravity's rainbow, Eq.\eqref{modtemp} predicts that the temperature of all black rings is modified by the ratio $g(E)/f(E)$. Thus, the modified temperature of charged black rings take the form 
\begin{equation}
\label{modtemp2}
T'=T\frac{g(E)}{f(E)}=\frac{1}{4\pi R\lambda\cosh(\beta)} \sqrt{1-\eta\left(\frac{1}{R\lambda E_P}\right)^n},
\end{equation}
where, following the same argument as before, we used 
\begin{equation}
E\geq \frac{1}{\Delta x}\approx\frac{1}{R\lambda}.
\end{equation}

\begin{figure}[h]
\centering
\includegraphics[width=0.5\linewidth]{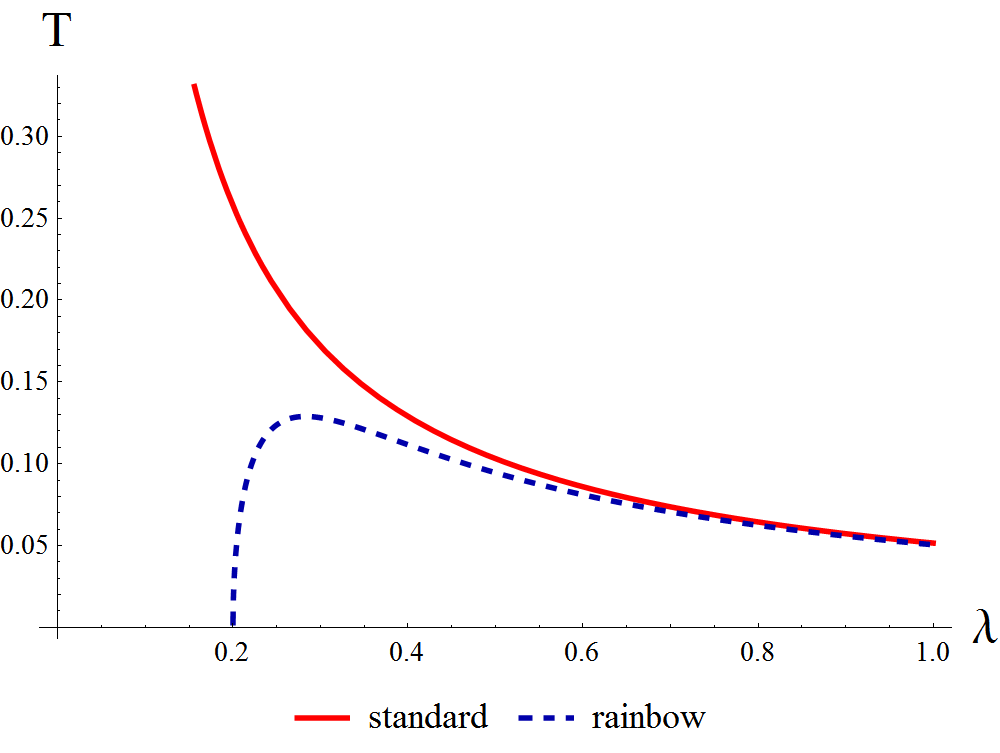}
\caption{\label{fig:temp2} Temperature of charged black rings as a function of $\lambda$.}
\end{figure}

Notice that the temperature goes to zero at $\lambda=\eta^{\frac{1}{n}}/RE_P$, which means that the mass has the minimum value
\begin{equation}
M_{min}=\frac{3}{4}\pi R\frac{\eta^{\frac{1}{n}}}{E_P}\left(1+\frac{\eta^{\frac{1}{n}}}{E_PR}\right)\left(1+\frac{2}{3}\sinh^2\beta\right).
\end{equation}
This implies that the charged black ring also has a remnant. Fig. \ref{fig:temp2} is a plot of the modified temperature with the generic values $\beta=1, E_P=5, n=2, \eta=1,$ and $R=1$.

\begin{figure}[h]
\centering
\includegraphics[width=0.5\linewidth]{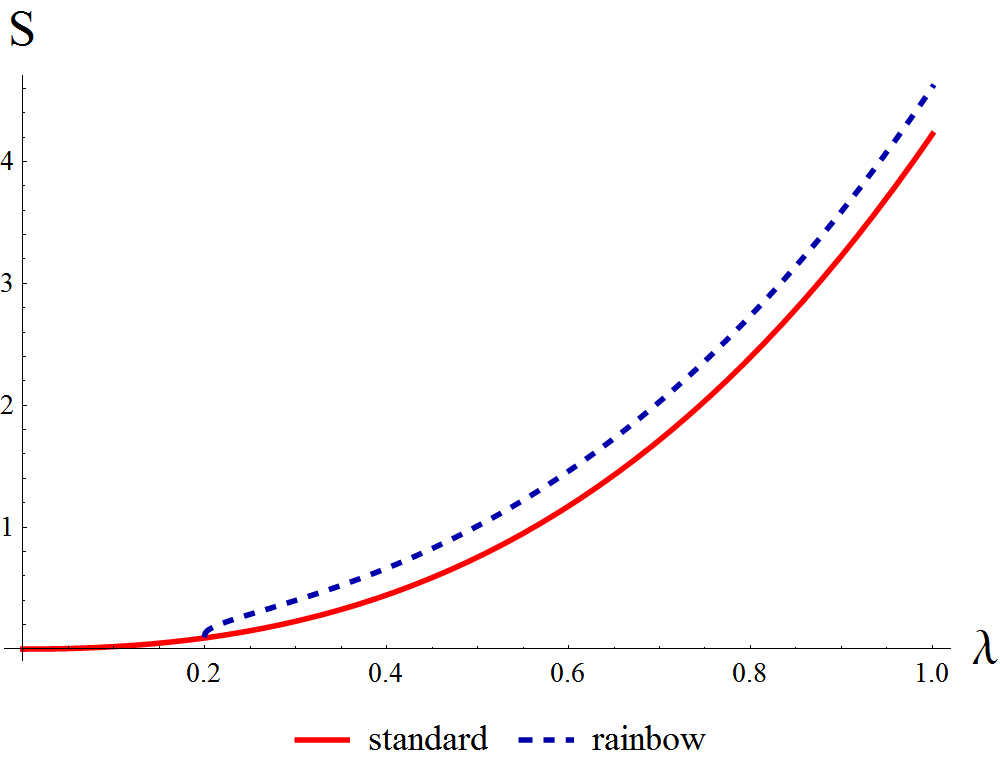}
\caption{\label{fig:ent2} Entropy of charged black rings as a function of $\lambda$.}
\end{figure}

The modified entropy becomes
\begin{equation}
S'=\int \pi^2R^3\lambda(1+2\lambda)\cosh(\beta)\left[4-\cosh(2\beta)\right] \frac{1}{\sqrt{1-\eta\left(\frac{1}{R\lambda E_P}\right)^n}} d\lambda
\end{equation}
using as a specific example $n=2$ we get
\begin{eqnarray}
S'=&&\frac{\pi^2\cosh\beta\left[4-\cosh(2\beta)\right]}{6R E_P^4\lambda\sqrt{1-\eta/(RE_P\lambda)^2}} 
\left[R^4 E_P^4\lambda^3(3+4\lambda)-8\eta^2+R^2 E_P^2\lambda\eta(4\lambda-3)
\right. \nonumber\\
&& \left.
+3RE_P\eta \sqrt{(R E_P\lambda)^2-\eta}\ln\left(R^2 E_P^2\lambda+R E_P\sqrt{(R E_P\lambda)^2-\eta}\right)\right]
\end{eqnarray}
The modified entropy is plotted in fig. \ref{fig:ent2}, and we clearly see it goes to zero at $\lambda=\eta^{\frac{1}{n}}/RE_P$.

\begin{figure}[h]
\centering
\includegraphics[width=0.5\linewidth]{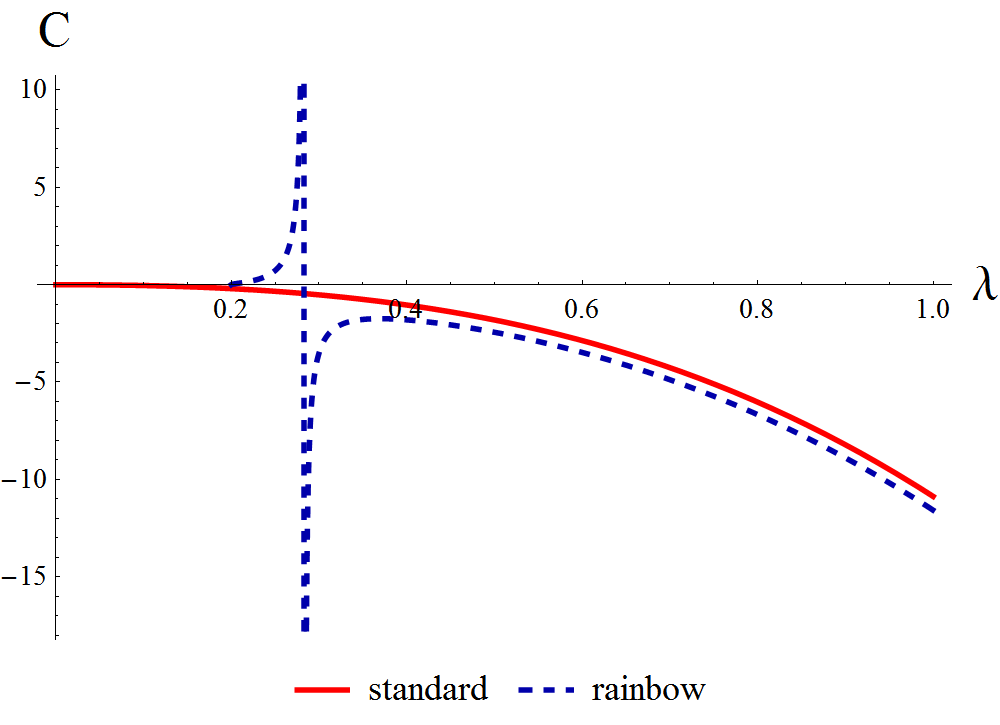}
\caption{\label{fig:capacity2} Heat capacity of charged black rings as a function of $\lambda$.}
\end{figure}

The modified heat capacity takes the form
\begin{equation}
C=\frac{\pi^2R^3\lambda^2(1+2\lambda)\left(25\lambda^2-1\right)\cosh(\beta)\left(\cosh(2\beta)-4\right)}{(25\lambda^2-2)\sqrt{1-\frac{\eta}{E_P^2R^2\lambda^2}}}
\end{equation}
From Fig. \ref{fig:capacity2} of the heat capacity, we see that the black ring is thermodynamically unstable, because $C<0$, contrary to the case of spinning black ring. However, for both cases the heat capacity diverges at maximum temperature and goes to zero at $\lambda=\eta^{\frac{1}{n}}/RE_P$, signaling the existence of a remnant.

\section{Conclusions}
We have analyzed a spinning black ring and a charged black ring  in gravity's rainbow and demonstrated that the rainbow functions change their  thermodynamic properties.
We have explicitly calculated corrections to the temperature,  entropy and heat capacity of the black rings. 
It was observed that the behavior of Hawking radiation changes considerably near the Planck scale in  gravity's rainbow. A direct consequence of this modification is that in the new  picture black rings do not evaporate completely; a black ring  remnant is left as the black ring evaporates down to Planck scale. 

It had been previously seen that such a remnant also exists for Schwarzschild  black holes in gravity's rainbow \cite{Ali:2014xqa}. It will be interesting to analyze other black objects like black saturns and AdS black holes in gravity's rainbow. We expect that the behavior we observed for black rings, (which was previously observed for the Schwarzschild black hole), will also occur for other black objects. This is because this behavior is generated from the rainbow functions, and using similar rainbow functions should produce the same effect for various black objects. However, it will be interesting to verify this for other black objects. 

\subsection*{Acknowledgments}
The research of AFA is supported by Benha University (www.bu.edu.eg) and CFP in Zewail City.

\bibliography{BR-rainbow}
 \bibliographystyle{JHEP}

\end{document}